\documentclass[pdflatex,sn-mathphys-num]{sn-jnl}

\usepackage{graphicx}%
\usepackage{multirow}%
\usepackage{amsmath,amssymb,amsfonts}%
\usepackage{amsthm}%
\usepackage{mathrsfs}%
\usepackage[title]{appendix}%
\usepackage{xcolor}%
\usepackage{textcomp}%
\usepackage{manyfoot}%
\usepackage{booktabs}%
\usepackage{algorithm}%
\usepackage{algorithmicx}%
\usepackage{algpseudocode}%
\usepackage{listings}%

\newcommand{\tool}{\textsc{SPVR}\xspace}

\newcommand{\nameofbaseline}{Sole LLM}

\usepackage{lineno}
\usepackage{hyperref}

\usepackage{babel}
\usepackage[font=small,labelfont=bf,tableposition=top]{caption}
\usepackage{booktabs}
\usepackage{algorithm}  
\usepackage{algorithmicx}  
\usepackage{algpseudocode}  
\usepackage{amsmath}  
\usepackage{pdfpages}
\usepackage{xspace}
\usepackage{pgf-pie}
\usepackage{threeparttable}
\usepackage{endnotes}
\usepackage{amssymb}
\usepackage{makecell}

\usepackage{tabularx}
\usepackage{array}

\usepackage{float}                 
\usepackage{overpic} 
\usepackage{caption,subcaption}
\usepackage{caption}
\usepackage{subcaption}

\usepackage{makecell}
\usepackage{tikz}
\usetikzlibrary{shapes.geometric}
\usepackage{multirow}

\usepackage{tablefootnote}

\usepackage{tcolorbox}
\usepackage{colortbl}
\usepackage{color, xcolor}
\usepackage{soul}

\theoremstyle{thmstyleone}%
\theoremstyle{thmstyletwo}%

\theoremstyle{thmstylethree}%

\newcommand{\wrk}[1]{#1}

\usepackage{lineno}
\usepackage{hyperref}

\usepackage{babel}
\usepackage[font=small,labelfont=bf,tableposition=top]{caption}
\usepackage{booktabs}
\usepackage{algorithm}  
\usepackage{algorithmicx}  
\usepackage{algpseudocode}  
\usepackage{amsmath}  
\usepackage{pdfpages}
\usepackage{xspace}
\usepackage{pgf-pie}
\usepackage{threeparttable}
\usepackage{endnotes}
\usepackage{amssymb}
\usepackage{makecell}

\usepackage{tabularx}
\usepackage{array}

\usepackage{float}                 
\usepackage{overpic} 
\usepackage{caption,subcaption}
\usepackage{caption}
\usepackage{subcaption}

\usepackage{makecell}
\usepackage{tikz}
\usetikzlibrary{shapes.geometric}
\usepackage{multirow}

\usepackage{tablefootnote}

\usepackage{tcolorbox}
\usepackage{colortbl}
\usepackage{color, xcolor}
\usepackage{soul}

\newcommand{\parh}[1]{\noindent\textbf{#1}}

\begin{document}

\title{SPVR: Syntax-to-Prompt Vulnerability Repair Based on Large Language Models}


\author[1]{\fnm{Ruoke} \sur{Wang}}\email{24S151159@stu.hit.edu.cn}

\author[2]{\fnm{Zongjie} \sur{Li}}\email{zligo@cse.ust.hk}

\author*[1]{\fnm{Cuiyun} \sur{Gao}}\email{gaocuiyun@hit.edu.cn}

\author[3]{\fnm{Chaozheng} \sur{Wang}}\email{adf111178@gmail.com}

\author[4]{\fnm{Yang} \sur{Xiao}}\email{xiaoyang@iie.ac.cn}

\author[1]{\fnm{Xuan} \sur{Wang}}\email{wangxuan@cs.hitsz.edu.cn}




\affil[1]{\orgname{Harbin Institute of Technology, Shenzhen},
\orgaddress{ \state{Shenzhen}, \country{China}}}

\affil[2]{\orgname{Hong Kong University of Science and Technology},
\orgaddress{ \state{Hong Kong}, \country{China}}}
\affil[3]{\orgname{The Chinese University of Hong Kong},
\orgaddress{ \state{Hong Kong}, \country{China}}}
\affil[4]{\orgname{Chinese Academy of Sciences},
\orgaddress{ \state{Beijing}, \country{China}}}






\abstract{\textbf{Purpose:} In the field of vulnerability repair, previous research has leveraged pre-trained models and LLM-based prompt engineering, among which LLM-based approaches show better generalizability and achieve the best performance. However, the LLM-based approaches generally regard vulnerability repair as a sequence-to-sequence task, and do not explicitly capture the syntax patterns for different vulnerability types, leading to limited accuracy. We aim to create a method that ensures the specificity of prompts targeting vulnerable code while also leveraging the generative capabilities of Large Language Models.

\textbf{Methods:} We propose SPVR (Syntax-to-Prompt Vulnerability Repair), a novel framework that collects information from syntax trees, and generates corresponding prompts. Our method consists of three steps: rule design, prompt generation, and patch generation. In the rule design step, our method parses code patches and designs rules to extract relevant contextual information. These rules aid in identifying vulnerability-related issues. In the prompt generation step, our method extracts information from vulnerable code with pre-defined rules, automatically converting them into prompts. We also incorporate the description of CWE (Common Weakness Enumeration) as known information into the prompts. Finally, in the patch generation step, this prompt will serve as input to any conversational LLM to obtain code patches.

\textbf{Results:} Extensive experiments validate that our method achieves excellent results in assisting LLMs to fix vulnerabilities accurately. We utilize multiple Large Language Models to validate the effectiveness of our work, repairing 143 of 547 vulnerable code using ChatGPT-4. We conducted a comparison of our approach against several existing vulnerability repair approaches (including fine-tuning-based and prompt-based), across multiple metrics. 

\textbf{Conclusion:} Our method is a novel framework that combines the Abstract Syntax Tree structure of code, providing targeted prompts of repair code for vulnerabilities. Our method demonstrates promising potential for real-world code vulnerability repair.}


\maketitle

\section{Introduction}
\label{sec:intro}

The rapid advancement of deep learning has transformed various domains, with Large Language Models (LLMs) achieving remarkable results in a wide range of tasks, often surpassing human performance in numerous natural language tasks~\cite{chen2021evaluating,ma2023oops,kasneci2023chatgpt,chang2024survey,kim2023language}. Leveraging the power of LLMs, researchers have successfully applied them to code-related fields, including code generation, code summarization, code translation, and code completion~\cite{
gu2021cradle, Jiang_2021}.
By seamlessly combining natural language and programming language, deep learning-based algorithms have demonstrated impressive performance in these areas~\cite{10.1145/3533767.3534390,liu2023exploring,
10172590,
wong2023binaug}. Building upon these advancements, the field of code vulnerability repair, a critical task in software engineering that enables developers to gain a deeper understanding of their code and improve the quality~\cite{wang2023reef,Wu_2023,fu2023learning,li2023split}, has also benefited from the application of neural networks to identify potential security flaws~\cite{10.1186/s13638-023-02255-2,li2023feasibility,wang2023instructta,fan2023automated,jiang2023impact}.

Existing approaches for code vulnerability repair using LLMs can be broadly categorized into two types. The first category focuses on 
fine-tuning models to learn code repair rules, thereby assisting in vulnerability repair~\cite{10.1145/3377811.3380338,vrepair,li2023vrptest}.
These approaches leverage the capacity of deep learning models to automatically identify and suggest fixes for potential security risks in codebases, streamlining the process of maintaining software security. 
The second category capitalizes on the powerful instruction-following abilities of LLMs, designing appropriate prompts to guide the model in repairing various types of vulnerabilities, thus enhancing their generalization capabilities~\cite{pearce2022examining}.
By extensively collecting vulnerabilities from commit patches or crafting suitable prompts, prior studies have proven the efficacy of LLMs in repairing code vulnerabilities~\cite{10.1145/3533767.3534387}.

In software systems, vulnerabilities frequently manifest through characteristic code patterns, which are systematically categorized by the Common Weakness Enumeration (CWE)~\cite{mitre2023Most}. For example, buffer overflows (CWE-120) often stem from inadequate boundary checks in C/C++ array operations; integer overflows (CWE-190) emerge from arithmetic computations lacking proper constraints; and memory management flaws (e.g., CWE-416: Use After Free) typically arise in contexts involving intricate pointer manipulations. Prior work has argued that treating vulnerabilities merely at the level of abstract CWE labels, while neglecting their underlying syntactic and semantic code structures, risks obscuring the essential correlations between taxonomy and implementation. Bridging this gap is thus pivotal for developing repair strategies that are both effective and generalizable~\cite{honkaranta2021towards,albanese2024cve2cwe, saletta2020neural,piran2021vulnerability}.

Software systems are inherently complex, often consisting of millions of lines of code across multiple languages, libraries, and frameworks~\cite{antinyan2017evaluating,
nunez2017source,shin2008empirical,datseris2024agents,idrisov2024program}. 
Such complexity gives rise to non-trivial interactions among variables, functions, and modules, which significantly increase the likelihood of introducing vulnerabilities~\cite{tambon2025bugs,lipp2022empirical,sakharkar2023systematic,wu2024knowledge}. Prior studies have shown that code complexity metrics (e.g., cyclomatic complexity, control-flow depth, and code churn) are positively correlated with defect-proneness and security risks~\cite{tehrani2024assessing,seyam2021code,stavtsev2024evaluating,jing2022improvement}. 
This underscores that vulnerability repair is not merely about fixing isolated syntax or semantic issues, but rather requires understanding the broader context in which vulnerabilities are embedded.

However, current approaches have two main limitations. 
First, due to the complexity of real-world code, vulnerability repairing issues can be challenging. Repairing these vulnerabilities often requires a deep understanding of both the specific programming language and the intricate interactions within the codebase~\cite{vrepair,zhang2022examplebasedvulnerabilitydetectionrepair,xin2019better,alarcon2020would,zhao2024repair}.
Second, existing works typically treat source code and vulnerability types separately in vulnerability repair. Some approaches solely focus on vulnerability type information and develop corresponding rules for handling them~\cite{alrashedy2024llmspatchsecurityissues}, while others incorporate additional code structure information such as Abstract Syntax Trees (AST)~\cite{nowack2021expanding}, Control Flow Graphs (CFG)~\cite{10.1145/3597926.3598142,Huo_Li_Zhou_2020}, or Program Dependence Graphs (PDG)~\cite{10.1145/2813885.2737957,noda2021sirius} to aid model learning. 
However, these segregated treatments hinder the model's ability to fully exploit the correlations between code and vulnerability types, potentially impacting its performance.

To address these limitations, we propose a novel framework that combines the AST structure from vulnerability and fix commits with CWE IDs to provide a more precise repair context and targeted instantiation of repair code for 
code vulnerability. Specifically, we utilize the widely adopted CWE~\cite {mitre2023Most} classification standard in the security community, along with our customized rules for identifying vulnerable code patterns based on the Minimal Edit Tree (MET) concept. By providing this information as contextual input to the model, along with specific modifications to the code under analysis, we enhance the model's generalization ability and improve vulnerability repair performance.

Our tool offers an up-to-date solution to assist the community in better leveraging LLMs for code vulnerability repair. Through the analysis of 547 code snippets, we summarize six common Minimum Edit Tree types and design their corresponding repair templates and rules. Extensive experiments demonstrate that our approach achieves excellent results on five popular LLMs, including Magicoder~\cite{wei2023magicoder}, Deepseek Coder~\cite{deepseek-coder}, ChatGPT-3.5~\cite{chatgpt}, ChatGPT-4, and Gemini~\cite{blogGemini25}. Notably, our method achieves a 26\% higher accuracy compared to an existing LLM-based 
vulnerability repair method, validating its effectiveness. Furthermore, our approach is independent of any specific LLM and can quickly adapt to new MET types, which is crucial in real-world scenarios where vulnerability categories and repair templates continuously evolve.

In summary, we make the following contributions:

\begin{itemize}
\item We present a novel abstract syntax tree, named \textbf{the Minimum Edit Tree (MET)}, to capture the most granular modifications made for repairs.
It is derived from modification records on the syntax tree based on vulnerability and fix commits, enabling a clearer identification to repair vulnerabilities.
We record the type of the MET's root node as MET type for subsequent analysis.

\item We establish a comprehensive set of rules based on MET types to guide the vulnerability repair process, facilitating the analysis of vulnerable code's MET and generating tailored prompts for LLMs. This approach leverages the advanced language understanding capabilities of LLMs to enable more nuanced and context-aware repairs.

\item We propose a novel framework that combines the MET of code with CWE information, providing a more precise repair context and targeted instantiation of repair code.
Our framework is independent of any specific LLM or LLM workflow and can rapidly adapt to new MET types. It shows promising applicability for practical application in real-world code vulnerability repair.

\end{itemize}
\section{Preliminary}
\label{sec:motivation}


This section introduces the background of the study, along with the motivation and challenges of the study.
\subsection{Background}
\parh{Program Analysis and Rule-based Methods.}
Program analysis and rule-based methods have been widely used for code vulnerability detection and repair. These methods rely on pre-defined rules or patterns to identify and fix improper operations within the source code. For example, Saber~\cite{10.1145/2338965.2336784} proposes symbolic rules to simulate programming logic and detect vulnerabilities. However, creating well-defined vulnerability rules or patterns heavily relies on expert knowledge~\cite{du2020leopard}, making it challenging to cover all potential cases. Moreover, the complex programming logic inherent in real-world software projects hinders the manual identification of these rules, limiting the performance of traditional 
program analysis and 
rule-based approaches in vulnerability repair~\cite{macedo2016feature,gao2024sguard+,rodriguez2022suggesting}.

\parh{Pre-trained Model-based Methods.}
Pre-trained model-based methods have emerged as a promising approach for vulnerability repair~\cite{10232867,9833850,10.1145/3643651.3659892}. These methods leverage the knowledge encapsulated in pre-trained models, such as CodeBERT~\cite{feng2020codebert}, CodeT5~\cite{wang2023codet5plus}, and UniXcoder~\cite{guo2022unixcoder}, which are designed to support both code-related classification and generation tasks. These approaches have achieved state-of-the-art performance by fine-tuning these pre-trained models on specific downstream tasks, such as vulnerability detection and repair. 
For instance, EPVD~\cite{10153647} proposes an execution path selection algorithm and adopts a pre-trained model to learn path representations for vulnerability repair.
SVulD~\cite{ni2023distinguishing} constructs contrastive paired instances and uses a pre-trained model to learn distinguishing semantic representations for code repair. 
These methods eliminate the need for expert involvement or the generation of structured graphs, making them more adaptable to real-world scenarios.


\parh{Prompt-based Vulnerability Repair Methods.}
Vulnerability repair is a complex and challenging task due to the intricacy of real-world code and the diversity of potential vulnerability types.
Recently, the rapid development of LLMs
, particularly the GPT series,
has simplified vulnerability repair to a certain extent, thanks to their strong generalization capabilities~\cite{tang2024code,yang2024multi,islam2024enhancing,yang2024cref}. By simply inputting the code snippet to be analyzed into the model, along with a specific instruction requesting the model to repair the code, LLMs can automatically provide repair suggestions (e.g., Figure \ref{motivation} (e)). The input can be a code snippet of arbitrary length, and the model will identify potential vulnerabilities within the given code based on the instruction and propose fixes.

\subsection{Motivation Example}
\begin{figure}[!htbp]
  \centering

    \includegraphics[page=1,width=0.85\textwidth]{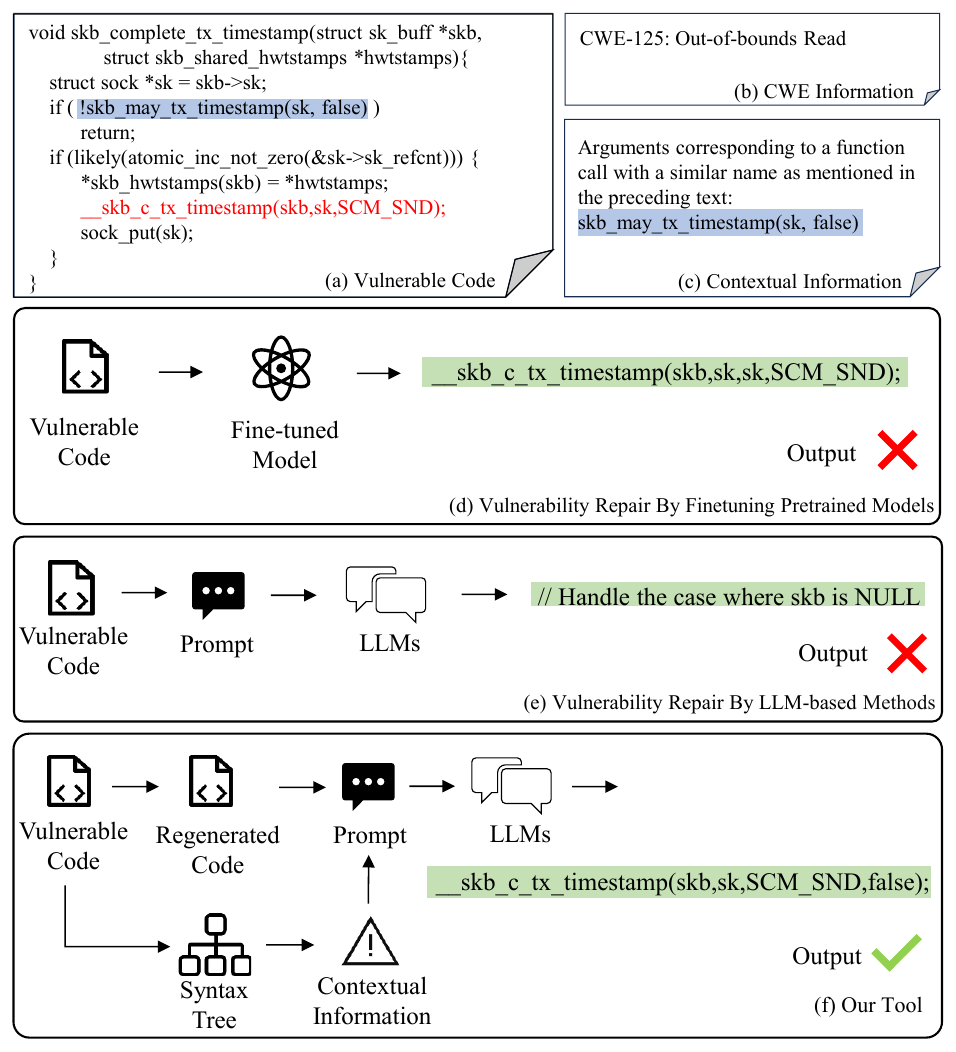}

  \caption{Motivation example demonstrating the limitations of existing repair approaches and the effectiveness of \tool.} 
  \label{motivation}

\end{figure}

Figure \ref{motivation} illustrates a motivating example, where a code snippet containing a vulnerability needs to be repaired.
In the vulnerable code snippet, Figure~\ref{motivation}(a) shows the function \texttt{skb\_complete\_tx\_timestamp}, where potentially relevant hints are highlighted with a blue background. Figure~\ref{motivation}(b) presents the associated vulnerability type, classified as CWE-125: Out-of-bounds Read. Figure~\ref{motivation}(c) provides the contextual information, namely the arguments of a function call with a similar name (\texttt{skb\_may\_tx\_timestamp(sk, false)}), which can serve as repair-relevant cues identified by \tool.

In Figure \ref{motivation} (d), a pre-trained model-based method is applied, while Figure \ref{motivation} (e) demonstrates a general LLM-based approach. However, both methods fail to resolve the issue effectively. 
The pre-trained model fails to understand the context of the code as well as the vulnerability type, leading to a few seemingly beneficial but useless changes at the syntax level. The general LLM-based approach, on the other hand, successfully identifies the error part but fails to provide a precise repair suggestion.
In contrast, our proposed method, shown in Figure \ref{motivation} (f), successfully addresses the problem. By parsing the code's syntax tree structure and providing it as additional information to the model, along with the regenerated code that highlights the potential error parts, LLMs can better understand the code and improve the accuracy of the repair process.

\subsection{Key Challenges}
Based on the above observations, we summarize the following key challenges:

\begin{itemize}
\item[1.] \textbf{Generalizability.} Pre-trained model-based methods, which rely on fine-tuning, have limited generalizability. Moreover, the constraints on input and output formats make it difficult for 
solving complex vulnerability repair problems.

\item[2.] \textbf{Contextual information extraction.} Prompt-based methods exhibit better generalizability but are influenced by the contextual information provided. Extracting relevant contextual information from the code to supply to the model remains a challenge.

\item[3.] \textbf{Linking code and vulnerabilities.} Establishing a connection between the code to be analyzed and the specific vulnerability types is another critical issue that needs to be addressed.
\end{itemize}

To tackle these challenges, we aim to develop a more effective and efficient approach for code vulnerability repair using LLMs, which can better understand the code structure and context, leading to improved repair accuracy and generalizability.





\section{Approach}
\label{sec:approach}

\begin{figure*}[!htbp]
    \centering
    \includegraphics[width = 1.00\textwidth]{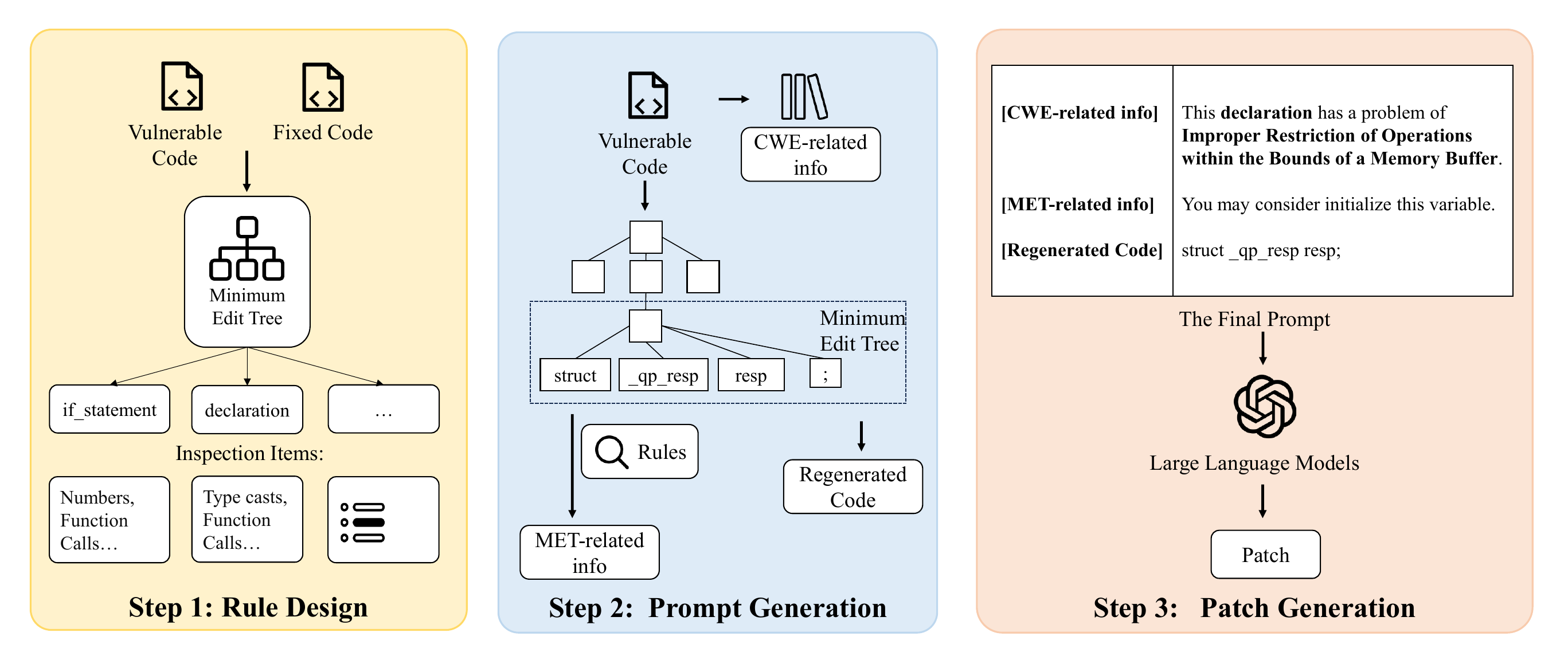}
    \caption{The overview of \tool.}
    \label{fig:overview}
\end{figure*}

This section details the methodology of our automated code vulnerability repair tool, \tool. Our approach is designed to effectively identify and fix code vulnerabilities by leveraging LLMs with carefully designed prompts. The core of our method involves a three-stage process: Rule Design, where we analyze code modifications to establish rules for collecting relevant contextual information; Prompt Generation, which uses these rules to create targeted prompts for LLMs; and Patch Generation, where we use the LLMs to generate and validate the final code patches.

\subsection{Overview}

We have developed a tool called \tool for code vulnerability repair. As shown in Figure \ref{fig:overview}, the tool consists of three steps: Rule design, prompt generation, and patch generation.

\parh{Rule design.}  
Given a pair of vulnerable code and its corresponding fixed version, we first identify the smallest unit of modification by comparing the two, and then derive the Minimum Edit Tree (MET). The MET captures fine-grained structural changes. On top of these fine-grained edits, we design a set of more abstract and coarse-grained rules that focus on high-level semantic aspects. These rules determine which contextual elements should be extracted, serving as the foundation for transforming low-level edits into meaningful, repair-relevant cues.

\parh{Prompt generation.}  
Different from the offline rule design stage where both vulnerable and fixed code are available, the online stage only has access to the vulnerable code and the location of the detected vulnerability. In this setting, we leverage external information to annotate the vulnerability type. We then revisit the MET and apply the rules defined in Step 1 to automatically extract MET-related information, such as critical variables or contextual function arguments. We regenerate code fragments to improve clarity, retaining only condition-related variables, key literals, return statements, or essential blocks, so that the prompt concisely integrates CWE information, MET-derived cues, and error-focused snippets.

\parh{Patch generation.}  
Finally, we integrate the three components into a prompt. This prompt explicitly communicates both the vulnerability type and the structural repair cues to an LLM. The LLM then uses this enriched context to generate a candidate patch. By combining precise static analysis signals with the generative capability of LLMs, \tool produces repair suggestions that are both syntactically correct and semantically aligned with the intended fix.

\subsection{Rule Design}

\textbf{Definition:} A \textbf{Minimum Edit Tree (MET)} is merged from subtrees of vulnerable and fixed code. It contains all modified nodes, and the type of its root node should be a subtype of an expression or a statement. It identifies the minimum portion of code modified when making a code change.
The \textbf{MET Type} refers to the type of a Minimum Edit Tree's root node. 

In the step of rule design, we complete the task of generating the MET, and design rules for different MET types.

\subsubsection{Generation of the MET}

The purpose of defining the MET is to distinguish between the inner context and the outer context. This distinction is crucial because code snippets are often lengthy, but in reality, only a small portion of the code is modified during the repair process. By locating the MET, we can focus on the core part of the code that needs to be fixed.

Additionally, separating the inner and outer contexts allows for a more targeted analysis, as the two contexts are analyzed differently. MET-related information largely originates from within the MET, with a minor portion coming from the context. Generally, we capture evidence that may cause a vulnerability inside the MET and look for variables for further use in the outer context.

\begin{algorithm} [!htbp]
\small
\renewcommand{\algorithmicrequire}{\textbf{Input:}}  
\renewcommand{\algorithmicensure}{\textbf{Output:}}
    \caption{The Generation of Minimum Edit Tree} 
    \label{alg_find_minimun_edit_node} 
    \begin{algorithmic}[1]

        

        \Require ASTs of vul \& fixed code, $v\_tree$, $f\_tree$. 
        
Code edited in vul \& fixed code, $v\_edit$, $f\_edit$. 
            
        \Ensure $minimum\_edit\_tree$

        \Function{GetMinimumTree}{$node, edit$}

            \State $any\_child\_contain \leftarrow False$
            \For{$child \in node.children$}{

                \If{$node.contains(child)$}
                
                    \State $any\_child\_contain \leftarrow True$
                
                \EndIf
            
            }

            \EndFor
        
            \If{( (isSubtypeOf($node$,  $'\_expression'$)
            
            or isSubtypeOf($node$,  $'\_statement'$))
            
            and contains($node, edit$) 
            
            and not $any\_child\_contain$)}
            
            \State \Return{$node$}

            \EndIf

            \For{$child \in node.children$}
            
                \State $answer$ $\leftarrow$ \Call{GetMinimumTree}{$child, edit$}
                
                \If{$answer \neq null$}
                
                    \State \Return{$answer$}
                
                \EndIf
            
            \EndFor
            
    \State \Return{$null$}

    \EndFunction



    \State$v\_min\_node \leftarrow$ \Call{GetMinimumTree}{$v\_tree.root, v\_edit$}
    \State$f\_min\_node \leftarrow$ \Call{GetMinimumTree}{$f\_tree.root, f\_edit$}

    \If{$v\_min\_node.type \neq f\_min\_node.type$}
        \State $minimum\_edit\_tree$ $\leftarrow$ $null$
    \Else
        \State $minimum\_edit\_tree  \leftarrow \newline$ \Call{MergeTrees}{v\_min\_node,f\_min\_node}
    \EndIf

    \end{algorithmic}

\end{algorithm}

Algorithm \ref{alg_find_minimun_edit_node} shows the method to generate the MET by traversing the syntax trees of the vulnerable and fixed code. Lines 1-18 define the main function \texttt{GetMinimumTree}. It checks if the current node satisfies three conditions: being a subtype of expression or statement, containing the edited code, and none of its child nodes containing the edited code (lines 2-10). If these conditions are met, the current node is considered the root node of vulnerable or fixed edit tree; otherwise, the function recursively calls itself (lines 11-16).

Lines 19-20 call the \texttt{GetMinimumTree} function to find the MET for both the vulnerable and fixed code. 
After collecting vulnerable and fixed edit trees, we combine them into the MET.
We focus on cases where the types of vulnerable and fixed METs are the same, and do not consider scenarios of different MET types.

\subsubsection{Designing rules based on MET types}
\label{subsec:cci}

Having generated the MET, the next step is to design a set of customized rules based on MET types to collect different contextual information.
Each rule checks whether this code segment adheres to a specific property under its corresponding MET type. Appropriate instructions are provided in the MET-related info to fix the vulnerability.

\begin{table}
    \setlength{\tabcolsep}{15pt}
\caption{\wrk{Common inspection items categorized by MET type.}}
\small

\begin{tabular}{|c|c|c|c|c|c|c|}



\hline
\multicolumn{1}{|c|}{Check}   & IF         & ASS     & CALL       & DEC    & FOR        & DEF   \\ \hline
\makecell{Variable}    & \checkmark & \checkmark &            &            & \checkmark &             \\ \hline
\makecell{Number\\Literal}     & \checkmark & \checkmark &            & \checkmark & \checkmark &             \\ \hline
\makecell{Function\\Call}      & \checkmark & \checkmark &            & \checkmark &            &             \\ \hline
\makecell{Type\\Cast}          &            & \checkmark &            & \checkmark &            &             \\ \hline
\makecell{Type}               &            &            &            & \checkmark &            & \checkmark  \\ \hline
\makecell{Condition}           & \checkmark &            &            &            & \checkmark &             \\ \hline

\end{tabular}

    \centering

    \begin{tablenotes}[para, online]\footnotesize\smallskip%
    \def\tnote#1{\protect\TPToverlap{\TPTtagStyle{#1}}}%

The horizontal axis abbreviations represent MET types: \textit{IF} (\textit{if\_statement}), \textit{ASS} (\textit{assignment\_expression}), \textit{CALL} (\textit{call\_expression}), \textit{DEC} (\textit{declaration}), \textit{FOR} (\textit{for\_statement}), and \textit{DEF} (\textit{function\_definition}).
    \end{tablenotes}

\label{table:analysis}
\end{table}
    








Although the collection process is largely similar, different inspections often involve various aspects of the code. We first enumerate some common inspections as follows:


\parh{Possible Variable Use}:
This inspection rule identifies variables for possible inclusion inside the MET. It records all variables which are not inside the MET, to determine which variables should be included. When a member variable is absent from the MET, but its parent appears at least twice, its MET-related information would be included. Moreover, when the individual variables are absent from the MET but appear at least twice in the context, and a similar variable (string similarity $>$ 0.5) exists in the MET, its MET-related information will also be included.

\parh{Number Literal Check}: This rule is used to inspect number literals that may lead to vulnerabilities. If a number literal other than ${0,1}$ appears in the MET, MET-related information will include re-examining whether the numerical values of number literals in the vulnerable code are correct. This rule is related to array out-of-bounds issues.
This inspection is particularly crucial for identifying and preventing array out-of-bounds vulnerabilities.

\parh{Function Call Check}: This rule is used to inspect function calls. Specifically, a $call\_expression$ always includes a function call, and we conduct more detailed inspections within the $call\_expression$. If a function call exists in the MET, MET-related information will include modifying the function call's name or parameters.

\parh{Type Cast Check}: This rule is used to inspect type conversions. If a type cast exists in the MET, MET-related information will include re-examining whether the type cast is correct. Although C++ offers multiple ways of type casting, the prompt only allows standard type casts, e.g., \texttt{a=(int)b}.
%

\parh{Type Check}: This rule is used to inspect the types of numeric variables. If variables of type $int/long/double$ exist in the MET, MET-related information will include considering using unsigned numbers, type casting, or size\_t. This rule is related to issues such as integer overflow or underflow, precision loss, or data truncation.


\parh{Condition Check}: This rule is used to inspect conditions in conditional statements. If conditional statements in the MET contain the symbols: ${>, >=, <, <=}$, MET-related information will include re-examining whether the conditional operators are correct. This rule is related to security vulnerabilities associated with buffer overflows or injection attacks.

By collecting contextual information based on the MET type and the associated checks, we can establish a strong link between the vulnerability type and the relevant code context. This connection proves invaluable in the subsequent repair process, enabling more targeted and effective fixes for the identified vulnerabilities.

In addition to these common inspection items that appear across multiple MET types, there are also individual items tailored for different METs. Detailed introduction of these individual items is shown in Appendix \ref{secA1}. Besides, we have gathered the distribution of all MET types, and the detailed data can be found in Section \ref{sec:setup}. Common inspection items categorized by MET type are shown in Table~\ref{table:analysis}.

\subsection{Prompt Generation}

In the step of prompt generation, we obtain each component required to generate prompts separately. 
\subsubsection{CWE-related information}

Typically, a piece of vulnerability code in datasets records its CWE-ID, which indicates its vulnerability type. We exclusively utilize the top 25 CWEs of 2023\cite{mitre2023Most} along with their corresponding descriptions, following the format: This $MET\ type$ has a problem of $CWE\_description$.

\subsubsection{MET-related information}

Given a piece of vulnerable code and its associated details, we generate its MET. Subsequently, we analyze the code based on the rules derived from the previous step, producing informative repair instructions. To enhance the repair process further, we simplify the source code by removing extraneous information, resulting in a more focused regenerated code.




A vulnerable code snippet may align with multiple rules within its MET type. We sequentially examine each applicable rule, generating up to $N$ prompts per vulnerable code segment. We chose to set $N=3$, as further detailed in Section \ref{sec:setup}. Across these prompts, only the MET-related information varies, while the other two components remain constant.

\subsubsection{Code Regeneration}

\label{subsub:co}

Notably, the context regeneration strategies vary depending on the MET type.

For $if$ and $for$ Minimum Edit Trees, vulnerabilities typically occur in the condition expressions. Consequently, the code within the MET may contain numerous irrelevant code blocks that can interfere with LLMs and potentially lead to input lengths exceeding the maximum input length permitted by the models. To address this issue, we construct a set comprising all variables that appear in the conditional section and variables predicted in the Possible Variable Use check. 
For each statement within the code block, if it contains any variable from the aforementioned ``possible variable use" set, or includes a $string\_literal$, or is a $return\_statement$, it will be retained in the regenerated code.

String literals often contain natural language descriptions of the code's execution state, and return statements may include return values indicating success or failure (e.g., -1 often denotes failure). 
Therefore, retaining these statements helps LLMs understand the significance of the code block.
For $function\_definition$ METs, since a function's definition always appears at the beginning, we preserve the code from the front to the back. For the other three MET types, we start from the position of the MET and preserve the MET itself along with the subsequent code as the regenerated code.


\subsection{Patch Generation}

Finally, the assembled prompts are fed as inputs into LLMs to generate patches. Despite incorporating directives in the prompts for standardized output formats, unexpected occurrences may still arise in the LLMs' output, such as misplaced $\verb|`|\verb|`|\verb|`|$ or natural language explanations.
Therefore, in the process of parsing the output code, we identify the first subtree that matches the MET type and treat it as the final answer. This subtree is then turned into plain code and used to perform the evaluation step against the ground truth.

\section{Experimental Setup}
\label{sec:setup}
In this section, we introduce the dataset used, the composition of prompts, the types of LLMs employed, and the evaluation metrics utilized in our experiments.
\subsection{Studied Dataset}

In this work, we choose CVEFixes~\cite{bhandari2021:cvefixes} as the dataset for its wide usage in previous works on vulnerability detection and repair. It encompasses many vulnerabilities along with their corresponding CWE IDs. The dataset is collected from real-world scenarios, reflecting genuine vulnerability situations encountered in actual software systems. We focus on vulnerabilities of C/C++ code on this dataset.
According to the information displayed on the CWE website, we focus on the top 25 CWEs in our experiments. Although our research is not directly related to specific CWE types, selecting these top CWEs provides more data and enhances our findings' generality.

To address concerns about potential dataset bias, the rules used to guide \tool are derived from less than 5\% of the dataset and are high-level abstractions rather than instance-specific patterns. Despite being learned from this small subset, these rules generalize well, covering 61.39\% of the vulnerable code samples (six MET types in Table~\ref{tab:table-nodetype}), which helps prevent the prompts from merely memorizing dataset-specific artifacts.

We preprocess the dataset by removing all code comments and filtering out cases where two code segments are identical after comment removal.
After preprocessing, we conduct the statistical analysis to examine the distribution of MET types among the sample cases. The results show that a few MET types account for the majority of the samples, indicating the practicality of our approach. 
Specifically, 76.77\% of the samples have the same MET type for both vulnerable and fixed parts, and 61.39\% of all samples are distributed across 6 MET types. Details of the distributions of MET type are shown in Table~\ref{tab:table-nodetype}.
\begin{table}
\caption{Statistics of METs.}
\begin{tabular}{lll}
\hline
MET type & Quantity & Percentage       \\ \hline
if\_statement          & 180      & \textbf{20.20\%} \\
assignment\_expression & 68       & \textbf{7.63\%}  \\
call\_expression       & 96       & \textbf{10.77\%} \\
declaration            & 98       & \textbf{11.00\%} \\
for\_statement         & 33       & \textbf{3.70\%}  \\
function\_definition   & 72       & \textbf{8.08\%}  \\
others                 & 137      & \textbf{15.38\%} \\
of different types     & 207      & \textbf{23.23\%} \\ \hline\end{tabular}

\label{tab:table-nodetype}
\end{table}


After initial experiments, we defined the maximum number of prompts corresponding to each vulnerable code, $N$, to be 3. This choice not only balances the exploration of diverse repair strategies with computational feasibility, but also takes into account the cost-efficiency of querying large-scale proprietary models, which typically adopt usage-based pricing. By constraining $N$, our framework remains both practical and reproducible in real-world deployment scenarios.

\subsection{Employed LLMs and Implementations}
We employ five renowned LLMs.

\parh{ChatGPT-3.5 and ChatGPT-4}\cite{chatgpt} are language models developed by OpenAI.
They are capable of handling both natural language and programming language tasks. 

\parh{Gemini 2.5 Pro}\cite{blogGemini25} is Google's most advanced reasoning Gemini model, featuring a step-by-step reasoning to ensure enhanced accuracy and performance.

\parh{DeepSeek Coder}\cite{deepseek-coder} is a series of code language models trained from scratch on code and natural language. The evaluation on coding-related benchmarks demonstrates that DeepSeek-Coder-Base-33B outperforms existing open-source code LLMs~\cite{rozière2024code,zheng2023codegeex,li2023starcoder}.

\parh{Magicoder}\cite{wei2023magicoder} is a model family designed for generating low-bias and high-quality instruction code. Magicoder-S-DS-6.7B has been shown to outperform GPT-3.5-turbo-1106\cite{chatgpt} and Gemini Ultra\cite{deepmindGeminiGoogle} on the HumanEval benchmark \cite{chen2021evaluating}.

We utilize Python along with the Tree-sitter toolkit\cite{treesitterTreesitterxFF5CIntroduction} for analyzing C/C++ code. 
Tree-sitter is a widely used parser generator and incremental parsing library that constructs concrete syntax trees for source code and efficiently updates them as the code is edited. It is designed to be general enough to support different programming languages, fast enough to run interactively during code editing, and robust enough to handle incomplete or erroneous code.

We employ ChatGPT-3.5-turbo-0125 model and ChatGPT-4-turbo in our experiments. As ChatGPT and Gemini are closed-source models, we utilize the Python interface.

For open-source models, we locally employ DeepSeek-Coder-6.7B-Instruct and Magicoder-S-DS-6.7B.
Our experiments with them are conducted in an environment of Ubuntu 22.04.3 LTS with 62GB of memory and 32 cores.
The hyperparameters used for DeepSeek Coder are $top\_k=50$ and $top\_p=0.95$, while for Magicoder, we set $temperature=0.5$.

The number of prompts per sample is three, because fewer prompts may lead to inaccurate answers, and too many prompts may increase costs.

\subsection{Evaluation Metrics}

\parh{CodeBLEU}\cite{ren2020codebleu} considers the surface match similar to the original BLEU\cite{papineni-etal-2002-bleu}, and the grammatical correctness and the logic correctness, leveraging the abstract syntax tree and the data-flow structure.
CodeBLEU provides a measure of matching the machine-generated code and the reference code.

\parh{Tree edit distance} is a metric used to quantify the similarity between two tree structures by measuring the minimum number of operations required to transform one tree into another. These operations typically include insertion, deletion, and substitution of nodes. The tree edit distance is commonly used in various fields such as computer science, bioinformatics, and natural language processing.

\parh{Perfect patches} refers to the requirement that the predicted results are identical to the ground truth in terms of string representation. Since C/C++ is indentation-insensitive, we remove the redundant indentation (e.g., whitespace and blank lines) to make a fair evaluation.
Here we use the pass@k~\cite{kulal2019spoc} perfect patches. 
For each prompt, the model generates k code snippets. If at least one of the code snippets equals the ground truth, the model succeeded at this prompt in k samples. In pass@3, the aggregation across multiple attempts is performed by checking whether at least one of the three generated snippets per sample correctly repairs the vulnerability.

\section{Evaluation}
\label{sec:evaluation}

In this section, we present our experimental results with five research questions and their corresponding answers.

\parh{RQ1} What is the performance of \tool compared to other existing tools in repairing 
vulnerabilities?
\newline
\parh{RQ2} How does the accuracy of \tool vary across different MET types?
\newline
\parh{RQ3} How do the hyperparameters impact \tool's performance?
\newline
\parh{RQ4} How effective are the rules 
derived
from MET type context analysis in improving the accuracy of vulnerability repair?
\newline
\parh{RQ5} 
Under what circumstances can \tool effectively repair vulnerabilities?

Through these research questions, we aim to comprehensively evaluate the accuracy, effectiveness, and robustness of our tool in repairing C/C++ vulnerabilities.

\subsection{RQ1: Effectiveness of \tool}
\subsubsection{Comparison with Baseline Models}
To evaluate the effectiveness of our approach, we conduct experiments comparing our tool with \nameofbaseline, which only contains a simple ``This code has a vulnerability." and the regenerated code.
We utilize four language models: ChatGPT-3.5-turbo, ChatGPT-4-turbo, DeepSeek-Coder-6.7B-Instruct, and Magicoder-S-DS-6.7B, and assess their performance both with and without our approach.

Table \ref{rq1:detail} and Table \ref{rq1:detail2} present the results of our experiments. When using our tool, ChatGPT-4-turbo achieves the highest accuracy, successfully repairing 143 out of 547 vulnerabilities. And the accuracy of ChatGPT-3.5 is slightly lower, repairing 138 vulnerabilities. DeepSeek-Coder and Magicoder repair 87 and 74 vulnerabilities, respectively. In contrast, when using the baseline models without our approach, the performance is greatly lower.
Moreover, we conducted paired t-tests on the CodeBLEU scores, and the resulting p-values (all $< 0.05$) indicate that the improvements of \tool over the \nameofbaseline\xspace are statistically significant.

\begin{table*}[htbp]  
\centering
\begin{threeparttable}

\caption{Comparison of \tool with \nameofbaseline \space in Pass@3 and CodeBLEU. We mark the best results in \textbf{bold}. 
}
\label{rq1:detail}  
\small
\begin{tabular}{ccccc}
\toprule[1pt]
Model    & Method& Pass@3  & \makecell{Code\\BLEU}&  p-value\tnote{1}  \\ \hline
\multirow{2}{*}{ChatGPT-4}  & \nameofbaseline   & 65/547  & 0.3025 &   \multirow{2}{*}{$1.15\times10^{-11}$}                \\
    & \tool & \textbf{143/547} & \textbf{0.4109}  &             \\ \hline
\multirow{2}{*}{ChatGPT-3.5}  & \nameofbaseline   & 55/547  & 0.3205  & \multirow{2}{*}{$5.42\times10^{-14}$}            \\
    & \tool & \textbf{138/547} & \textbf{0.4456}  &                \\ \hline
\multirow{2}{*}{DeepSeek-Coder}& \nameofbaseline   & 32/547  & 0.2753  & \multirow{2}{*}{$8.93\times10^{-9}$}          \\
          & \tool & \textbf{87/547}  & \textbf{0.3889}  &                   \\ \hline
\multirow{2}{*}{Magicoder}    & \nameofbaseline   & 19/547  & 0.2795   & \multirow{2}{*}{$9.97\times10^{-10}$}            \\
      & \tool & \textbf{78/547}  & \textbf{0.3893    }   &             \\ \hline
  \multirow{2}{*}{Gemini-2.5-Pro}    & \nameofbaseline   & 66/547  & 0.3040   & \multirow{2}{*}{$4.80\times10^{-17}$}               \\
  & \tool & \textbf{136/547}  & \textbf{0.4566   }                \\

\bottomrule[1pt]
\end{tabular}

\begin{tablenotes}
        \footnotesize
        \centering
        \item[1]p-value is the t-test p-value for CodeBLEU. A p-value less than 0.05 indicates that the difference between \tool and \nameofbaseline \xspace is statistically significant.
        
\end{tablenotes}

\end{threeparttable}

\end{table*}

\begin{table}[!htbp]

\caption{Comparison of \tool with \nameofbaseline \space in four metrics of CodeBLEU. We mark the best results in \textbf{bold}. 
}
\centering
\small
\begin{tabular}{cccccc}
\toprule[1pt]
Model    & Method & \makecell{N-gram} & \makecell{Weighted\\N-gram} & \makecell{Syntax} & \makecell{Dataflow} \\ \hline
\multirow{2}{*}{ChatGPT-4}  & \nameofbaseline             & 0.2197             & 0.2353                        & 0.4346              & 0.3204                      \\
    & \tool           & \textbf{ 0.3233}              & \textbf{0.3441}                        & \textbf{0.5523}               & \textbf{0.4237 }                         \\ \hline
\multirow{2}{*}{ChatGPT-3.5}  & \nameofbaseline            & 0.2459              & 0.2616                        & 0.4771               & 0.2975                      \\
    & \tool           & \textbf{0.3694}              & \textbf{0.3839}                        & \textbf{0.6425}               & \textbf{0.3867}                         \\ \hline
\multirow{2}{*}{DeepSeek-Coder}& \nameofbaseline            & 0.2090              & 0.2219                        & 0.4292               & 0.2411                 \\
          & \tool           & \textbf{0.3110}              & \textbf{0.3173}                        & \textbf{0.5963 }              & \textbf{0.3310 }                \\ \hline
\multirow{2}{*}{Magicoder}    & \nameofbaseline            & 0.2097              & 0.2190                        & 0.4473               & 0.2419                \\
      & \tool    & \textbf{0.3154}              & \textbf{0.3231  }                      & \textbf{0.5846 }              & \textbf{0.3343}                  \\  \hline
\multirow{2}{*}{Gemini-2.5-Pro}    & \nameofbaseline            & 0.0937            & 0.2810                      & 0.3946              & 0.4468           \\
      & \tool    & \textbf{0.3515}              & \textbf{0.3009  }                      & \textbf{0.4663 }              & \textbf{0.7077}                  \\

\bottomrule[1pt]
\end{tabular}

\label{rq1:detail2}
\end{table}

The experimental results clearly demonstrate the effectiveness of our approach in enhancing the performance of vulnerability repair. By incorporating our methods, the accuracy of all four language models improves substantially. ChatGPT-4 and ChatGPT-3.5-turbo experience over 120\% increase in the number of successfully repaired vulnerabilities, while DeepSeek-Coder-6.7B-Instruct and Magicoder-S-DS-6.7B see improvements of 172\% and 311\%, respectively.

To further validate the reliability of these improvements, we collected an additional set of 300 new vulnerabilities published since 2024, following the data collection and preprocessing procedure of PatchDB \cite{wang2021PatchDB}. 
We used ChatGPT-4 for experiments on this new dataset. 
The results show that the Sole LLM successfully repaired 26 cases, whereas SPVR repaired 71 instances.
This consistent performance gap provides further evidence that the improvements achieved by SPVR are not merely due to possible data overlap, but rather from SPVR’s process of extracting syntax-based rules to generate targeted prompts that guide LLMs in producing effective vulnerability patches.

These findings highlight our approach's significance in augmenting LLMs' capabilities for vulnerability repair tasks. The integration of our methods, such as context-aware rule derivation and prompt engineering, enables the models to better understand and address the complexities of C/C++ vulnerabilities.
Besides, our approach is able to repair CWEs related to memory, such as CWE-119.
The consistent improvement across all three models demonstrates the generalizability and robustness of our approach.

\subsubsection{Comparison with Existing Tools}
To further evaluate the effectiveness of our tool, we compare its performance with existing vulnerability repair tools, namely VRepair\cite{vrepair} and RING\cite{joshi2022repair}.

VRepair is a fine-tuned model based on CodeT5+~\cite{wang2023codet5plus} that aims to repair vulnerabilities in C language functions. The evaluation metric of VRepair is perfect patches, where a repair is considered successful if the ground truth appears among the top 50 outputs 
(beam size = 50).

As shown in Table \ref{comaprevrepair}, our tool achieves a slightly higher accuracy of perfect patches than VRepair. The accuracy of \tool is 26.14\% compared to VRepair's 22.73\%. It is important to note that VRepair employs a beam size of 50 for output generation, while our approach uses pass@3 evaluation. Despite this difference, we retain the beam size of 50 in VRepair for a fair comparison, although a smaller beam size can be more practical in real-world scenarios.

As shown in Table \ref{tab:comaprering}, RING is an LLMC-based approach to multilingual repair that utilizes diagnostic messages, smart few-shot selection, and ranking of repair candidates. RING evaluates vulnerability repair in C language using the tree edit distance metric. A patch is considered successful if the tree edit distance between the output and the ground truth is less than 5.
When comparing the tree edit distance, our tool demonstrates significantly better performance than RING. The pass@3 proportion of edit distances $<$ 5 of our tool is 79.5\%, while RING's proportion is 63.0\%. This indicates that under the metric condition, the accuracy of \tool exceeds RING by 26.2\%. 




    



        
        
    


\begin{table}[!htbp]

    \centering
    
    \setlength{\tabcolsep}{15pt}
	\caption{Comparison with VRepair.}	

 	\begin{tabular}{c c}
		\toprule[1pt]
   Tool  & Perfect Patches(\%)\\
		\midrule 

  \tool & 26.14 \\
            VRepair   & 22.67 \\
		\bottomrule [1pt]
	\end{tabular}

    \label{comaprevrepair}
\end{table}

\begin{table}[!htbp]

\centering
        \caption{Comparison with RING.}	

        \begin{tabular}{c c c}
            \toprule[1pt]
             Tool & Pass@3(\%)(Edit distance $<$ 5)\\
            \midrule 
            \tool & 79.5 \\
            RING  & 62.8 \\
            \bottomrule [1pt]
        \end{tabular}

        \label{tab:comaprering}

\end{table}

\begin{figure}
    \centering
    \includegraphics[width=0.95\linewidth]{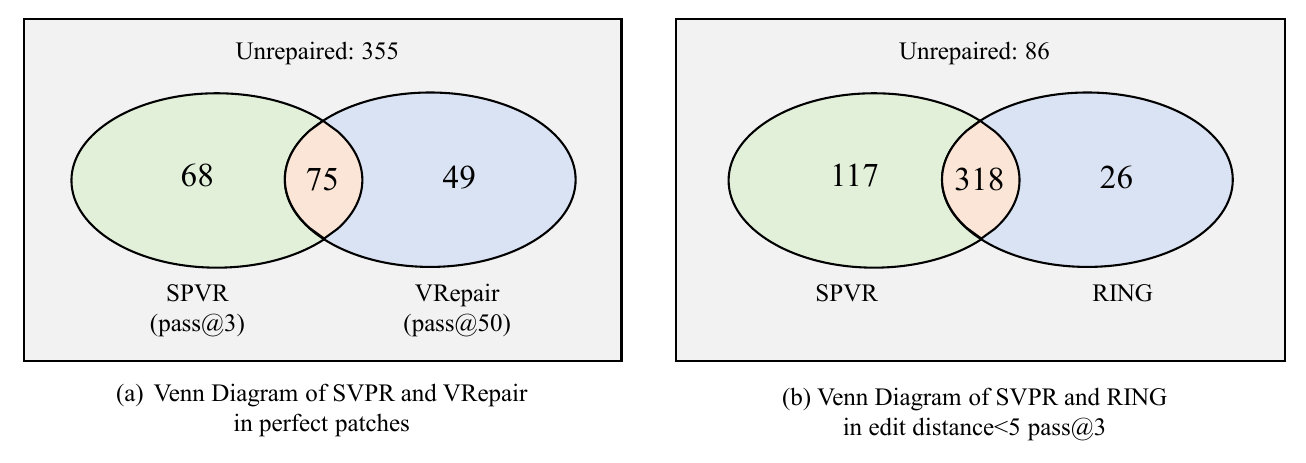}
    \caption{Venn Diagram for comparison \tool with VRepair and RING}
    \label{fig:venn}
\end{figure}

A Venn diagram of the experimental results is presented in Figure~\ref{fig:venn}. When compared to VRepair, \tool uniquely repairs 68 vulnerable code samples that VRepair cannot, while VRepair exclusively repairs 49 samples; both methods successfully repair 75 overlapping samples, and 355 samples remain unpatched by either approach. In comparison with RING, \tool independently repairs 117 samples, whereas RING alone fixes 26, with 318 overlapping repairs and 86 samples left unresolved.

The comparison with existing tools highlights the competitiveness and effectiveness of our approach. Despite the differences in evaluation metrics and experimental settings, our tool achieves superior performance compared to vulnerability repair tools like VRepair and RING.

\begin{tcolorbox}
\parh{Answer to RQ1:} The comparison with LLMs and traditional baselines strongly supports our tool's effectiveness in enhancing vulnerability repair performance. The improvements validate the value of our proposed methods and their potential to advance the state-of-the-art in automated vulnerability repair.
\end{tcolorbox}

\subsection{RQ2: Influence of Different MET Types}

To investigate the influence of different MET types on the performance of our tool, we conduct experiments to analyze the CodeBLEU scores and perfect patch rates (pass@3) for each MET type separately. Table \ref{rq2-1} and Table \ref{rq2-2} present the detailed results for five models.

\begin{table*}[!htbp]
\small
\begin{threeparttable}

\centering
\caption{Results of ChatGPT-3.5/4 and Gemini-2.5-Pro separated by MET type.}

\begin{tabular}{ccccccccc}
\toprule[1pt]

\scriptsize
                       &          & \multicolumn{2}{c}{ChatGPT-4}& \multicolumn{2}{c}{ChatGPT-3.5} &    \multicolumn{2}{c}{Gemini-2.5-Pro}   \\
MET type\tnote{1}                & Qty\tnote{2}  & \small{CodeBLEU  }         & Pass@3 & CodeBLEU           & Pass@3     & CodeBLEU           & Pass@3       \\ \hline
IF          & 180      &0.3600&41& 0.4531             & 41          &0.4272&   40                    \\
ASS & 68       &0.5088&13& 0.4656             & 9          &0.4326&         6          \\
CALL       & 96       &0.3274&25& 0.3726             & 17         &0.3976&      21                    \\
DEC            & 98       &0.3753&24& 0.3398             & 19        &0.3507&       17                \\
FOR         & 33       &0.6779&4& 0.6322             & 5        &0.6201&           1              \\
DEF   & 72       &0.5109&36& 0.5638             & 47          & 0.7005&              51         \\
\bottomrule[1pt]

\end{tabular}

\begin{tablenotes}
        \footnotesize
        \centering
        \item[1]The vertical axis abbreviations represent MET types: \textit{IF} (\textit{if\_statement}), \textit{ASS} (\textit{assignment\_expression}), \textit{CALL} (\textit{call\_expression}), \textit{DEC} (\textit{declaration}), \textit{FOR} (\textit{for\_statement}), and \textit{DEF} (\textit{function\_definition}).
        \item[2] Qty is the abbreviation of Quantity.

      \end{tablenotes}
\label{rq2-1}
\end{threeparttable}

\vspace{10pt}

\centering
\begin{threeparttable}

\centering
\caption{Results of DeepSeek-Coder and MagiCoder separated by MET type.}

\begin{tabular}{c c @{\hspace{30pt}} c c @{\hspace{20pt}} c c c}
\toprule[1pt]

\scriptsize
                       &      & \multicolumn{2}{c}{DeepSeek-Coder} & \multicolumn{2}{c}{MagiCoder} \\
MET type\tnote{1}                 & Qty\tnote{2}            & CodeBLEU                 & Pass@3                & CodeBLEU            & Pass@3            \\ \hline
IF                 & 180               & 0.3466                   & 18                    & 0.3494              & 15                \\
ASS          & 68               & 0.4512                   & 3                     & 0.4486              & 1                 \\
CALL                & 96              & 0.3065                   & 6                     & 0.3408              & 10                \\
DEC                       & 98               & 0.3548                   & 20                    & 0.3452              & 15                \\
FOR               & 33               & 0.5888                   & 4                     & 0.5585              & 1                 \\
DEF          & 72              & 0.5004                   & 36                    & 0.4813              & 32               \\
\bottomrule[1pt]

\end{tabular}

\begin{tablenotes}
        \footnotesize
        \centering
        \item[1]The vertical axis abbreviations represent MET types: \textit{IF} (\textit{if\_statement}), \textit{ASS} (\textit{assignment\_expression}), \textit{CALL} (\textit{call\_expression}), \textit{DEC} (\textit{declaration}), \textit{FOR} (\textit{for\_statement}), and \textit{DEF} (\textit{function\_definition}).
        \item[2] Qty is the abbreviation of Quantity.
        
      \end{tablenotes}
\label{rq2-2}
\end{threeparttable}

\end{table*}

The experimental results reveal that, although the total number of vulnerabilities fixed varies among different models, the proportion of successful repairs across MET types remains roughly the same. This observation suggests that there are inherent differences in the repair capabilities of our approach for different MET types, independent of the specific model used.

One possible explanation for these differences is that the rules summarized during the offline stage may not necessarily encompass all scenarios. Furthermore, when variables, constant names, or function names that never appear in the vulnerable code are added to the fixed code, it becomes challenging to generate accurate repairs. This is because our approach fundamentally relies on gathering existing information from the vulnerable code to provide prompts.

The scenario described above is particularly common when the MET type is either $if\_statement$ or $assignment\_expression$. For these MET types, the CodeBLEU scores of the generated outputs are relatively high.
However, the perfect patch rates (pass@3) for these MET types are notably low, suggesting that the generated code may not exactly match the ground truth, despite being close in terms of syntactic structure.
These findings underscore the importance of refinement of repair strategies tailored to specific MET types to enhance the effectiveness of vulnerability repair.

\begin{tcolorbox} 
\parh{Answer to RQ2:} The analysis of the influence of different MET types on repair performance reveals inherent variations in the effectiveness of our approach. The challenges of certain MET types highlight the limitations of relying solely on existing information from the vulnerable code.
%
\end{tcolorbox}

\subsection{RQ3: Impact of Hyperparameters}

To investigate the impact of hyperparameters on the output quality and accuracy of open-source models, we conduct experiments using DeepSeek-Coder-6.7B-Instruct. We vary the temperature parameter, which controls the randomness of the generated output, and evaluate its effect on the CodeBLEU scores and perfect patch rates (pass@3).

\begin{table}
\centering
\captionsetup{justification=centering, singlelinecheck=false}
\caption{DeepSeek Coder performance \\ with different temperatures.}
\centering

\begin{tabular}{cccc}
\toprule[1pt]
& \multicolumn{3}{c}{Temperature} \\
        & 0.25 & 0.5 & 0.75 \\ \hline
CodeBLEU & 0.3884 & 0.3889 & 0.3955 \\
Pass@3 & 86 & 87 & 91 \\
\bottomrule[1pt]

\end{tabular}

\label{rq3}
\end{table}

Table \ref{rq3} presents the results of our experiments with temperature values of 0.25, 0.5, and 0.75. The CodeBLEU scores for the generated outputs remain relatively stable across different temperature settings.
Similarly, the perfect patch rates (pass@3) do not exhibit significant changes, ranging from 86 to 91 successful repairs.

These results indicate that the model's ability to generate syntactically correct code and exact matches is not heavily influenced by the randomness introduced by temperature.

\begin{tcolorbox} 
\parh{Answer to RQ3:} The performance of \tool is not heavily dependent on the hyperparameter, providing confidence in its ability to generate reliable repairs under various settings.
\end{tcolorbox}

\begin{table}[t]

\centering
\captionsetup{justification=centering, singlelinecheck=false}
\caption{Result for ablation study with ChatGPT-3.5-turbo.}
\begin{tabular}{ccc}
\toprule[1pt]
                 & CodeBLEU & Pass@3 \\ \hline
Baseline (\tool) & 0.4456   & 138    \\
Without\_MET\_info     & 0.2646   & 55     \\
Without\_CWE\_info & 0.4380   & 128   \\
\bottomrule[1pt]
\end{tabular}

\label{rq4}

\end{table}

\subsection{RQ4: Ablation Study}
To gain a deeper understanding of the contribution of each component in \tool, we conduct an ablation study with ChatGPT-3.5-turbo-0125 model and same settings as RQ2. As described in section \ref{sec:setup}, our prompt consists of three components: information related to CWE, information related to MET, and regenerated code.
We do not replace regenerated code because LLMs often fail to locate the areas to be fixed, even when we append ``//fix this" at the end of the lines requiring modification in preliminary experiments.

We design three different combinations of these components to evaluate their impact on the output accuracy:

1. C1 (\tool): This 
is
our complete prompt, which includes all three components: CWE-related information, MET-related information, and the regenerated code.

2. C2 (without\_MET\_info): In this combination, we remove the MET-related information from the prompt, keeping only the CWE-related information and the regenerated code.

3. C3 (without\_CWE\_info): Here, we remove the CWE-related information from the prompt, retaining only the MET-related information and the regenerated code.

Table \ref{rq4} presents the results of our ablation study, comparing the CodeBLEU scores and pass@3 rates for each combination. The ablation study reveals several interesting findings. When CWE-related information is omitted (C3), the performance of our tool remains relatively high, with a pass@3 rate of 
128/547. The CodeBLEU score also remains close to the baseline (\tool), indicating that the generated code is syntactically similar to the ground truth.

In contrast, when MET-related information is removed from the prompt (C2), we observe a significant drop in performance. The pass@3 rate decreases to
55/547, representing a loss of over 50\% compared to the baseline. The CodeBLEU score also decreases substantially, indicating that the 
code deviates more from the ground truth. This highlights the importance of MET-related information in our prompt for achieving high repair accuracy.

The discrepancy in the impact of removing CWE-related and MET-related information can be attributed to the nature of the prompts.
The prompts related to MET types provide specific and direct instructions on repairing vulnerabilities based on the AST structure of the code.
On the other hand, the prompts related to CWE offer more abstract and high-level descriptions of the vulnerability types. The specificity and directness of the MET-related prompts appear to be more effective in guiding the model toward accurate repairs compared to 
the CWE-related prompts. This observation emphasizes the importance of incorporating fine-grained information in the prompts to effectively assist the model in generating precise repairs.

\begin{tcolorbox} 
\parh{Answer to RQ4:} Our ablation study demonstrates the contribution of MET-related information to the performance of our tool. 
This finding underscores the importance of incorporating syntax-aware prompts in vulnerability repair tasks to achieve high accuracy and effectiveness.
\end{tcolorbox}

\subsection{RQ5: Case Study}







\begin{figure}
  \centering
    \includegraphics[page=1,width=0.75\textwidth]{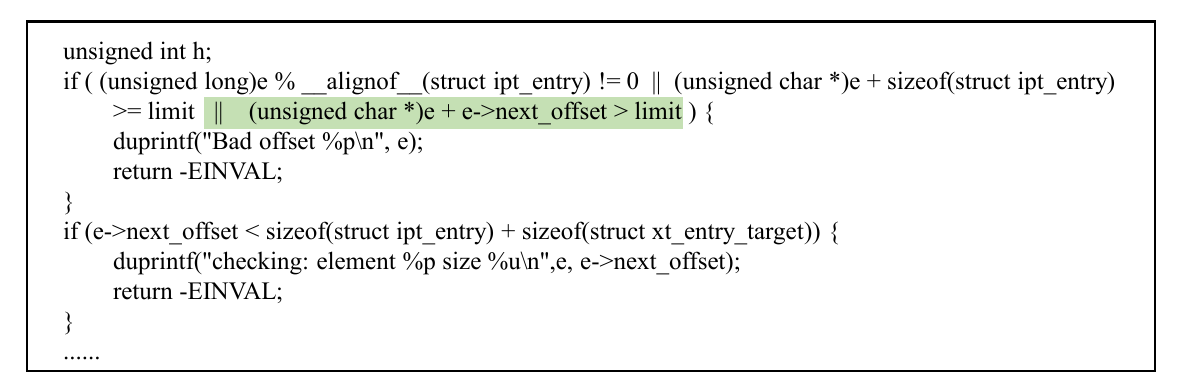}
  \caption{Case study: Vulnerable code and its patches.}
  \label{case1-1}
\end{figure}

In this section, we present a representative case study to examine the circumstances under which \tool can successfully repair vulnerabilities and to clarify the reasons for its effectiveness.

The case is shown in Figure~\ref{case1-1}. The CWE ID of the vulnerable code is CWE-119, i.e., Improper Restriction of Operations within the Bounds of a Memory Buffer. The correct fix adds an additional constraint in the \texttt{if} condition: \texttt{|| (unsigned char *)e + e->next\_offset > limit}.

\begin{figure}
  \centering
    \includegraphics[page=1,width=0.75\textwidth]{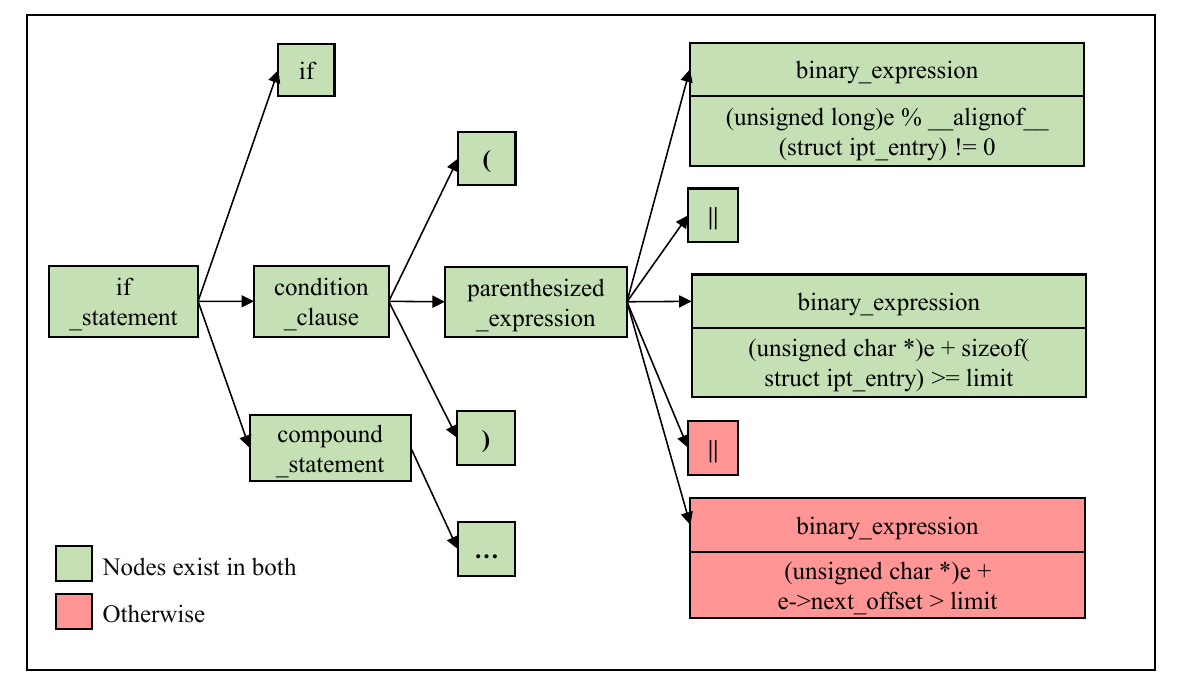}
  \caption{Case study: Vulnerable code and its Abstract Syntax Tree.}
  \label{case1-2}
\end{figure}

\begin{figure}
  \centering
    \includegraphics[page=1,width=0.75\textwidth]{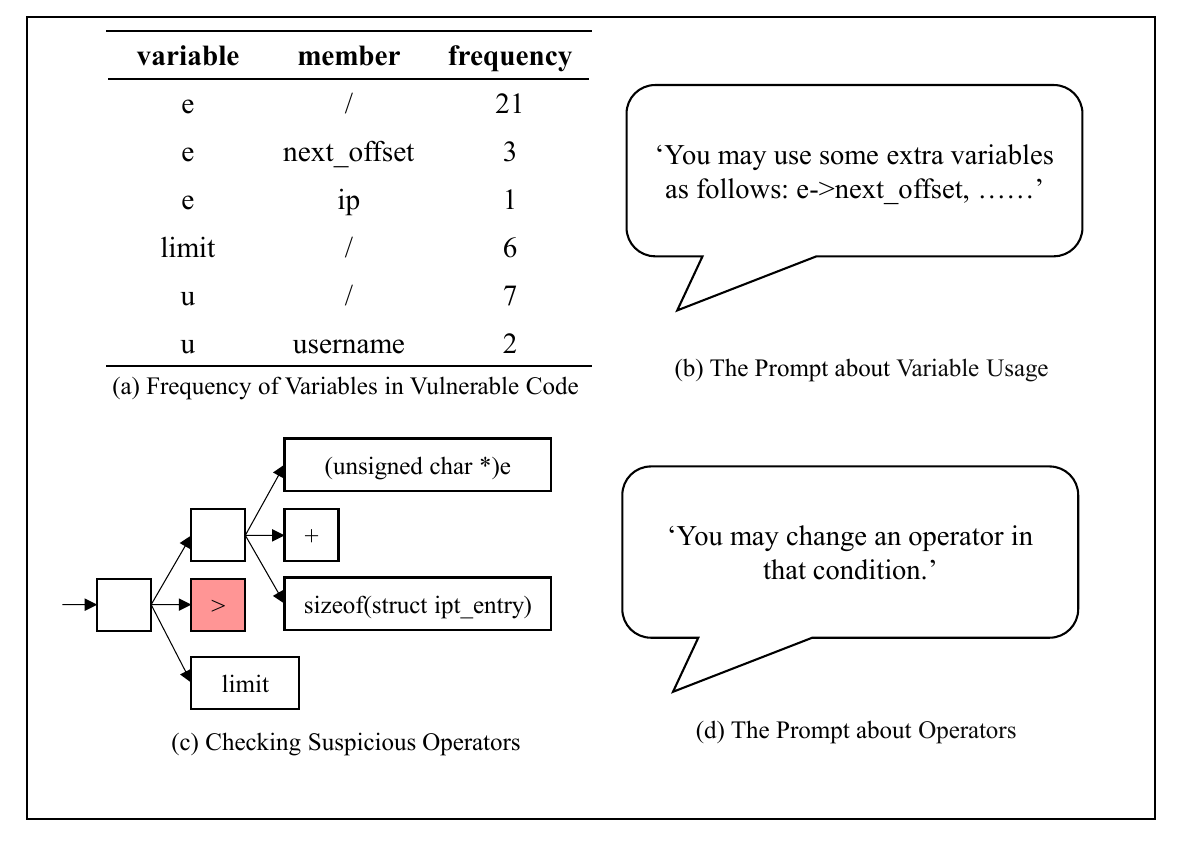}
  \caption{Case study: Analysis and prompt generation.}
  \label{case1-3}
\end{figure}

\tool analyzes the usage of variables in this code snippet through the Abstract Syntax Tree and counts the frequency of variables and their members in the surrounding context, as shown in Figure~\ref{case1-2}. The node names in the code tree follow the naming conventions of Tree-sitter.

Since the vulnerable node has the MET type \texttt{if\_statement}, we follow the ``IF” column in Table~\ref{table:analysis} to inspect up to three items, which are then summarized into three MET-related prompts.
During the inspection, the correct code in red is unknown; only the vulnerable code in green is available. 

After counting variable names, a member variable \texttt{e->next\_offset} is observed to occur frequently in the surrounding context but does not appear within the condition statement, as shown in Figure~\ref{case1-3}(a). This discrepancy suggests that the absence of this variable may be a potential cause of the vulnerability. Consequently, the first MET-related prompt explicitly recommends to the Large Language Model that \texttt{e->next\_offset} should be incorporated into the \texttt{if} condition, guiding LLMs to generate a fix that captures the missing semantic constraint, as illustrated in Figure~\ref{case1-3}(b).

Additionally, as Figure~\ref{case1-3}(c) shows, since the condition contains a binary operator, the vulnerability might also be caused by its usage. Accordingly, the second MET-related prompt encourages LLMs to examine the binary operator in the \texttt{if} condition, as shown in Figure~\ref{case1-3}(d).
Because the \texttt{if\_statement} may contain many irrelevant or redundant statements that could negatively affect the quality of patches generated by the LLM, the regenerated code retains only the statements involving the target variables, string literals, and return statements. The code regeneration part is consistent with Section~\ref{subsub:co}.

After generating up to three MET-related prompts, each of these prompts is combined with the CWE-related prompt and the regenerated code to form the final input for LLMs. Among the three responses to the first MET-related prompt, ChatGPT-3.5/4, Gemini-2.5-Pro, DeepSeek-Coder, and MagiCoder all produced the correct fix. Notably, even without the CWE-related prompt (as shown in the ablation study), ChatGPT-3.5 can still correctly repair this vulnerability. However, without the MET-related prompt, none of the LLMs above could fix the vulnerable code. 

This demonstrates the effectiveness of \tool, which performs semantic analysis of the vulnerable code and highlights potential repair strategies. At the same time, \tool leverages the strong comprehension and generative capabilities of LLMs. For example, the model may implicitly understand that \texttt{e->next\_offset} refers to a certain offset and that \texttt{limit} represents a boundary condition, thereby enabling it to generate the correct fix (even without the CWE-related prompt, which contains information that the vulnerability is about improper bounds of memory buffers).

\begin{figure}
  \centering
    \includegraphics[page=1,width=0.85\textwidth]{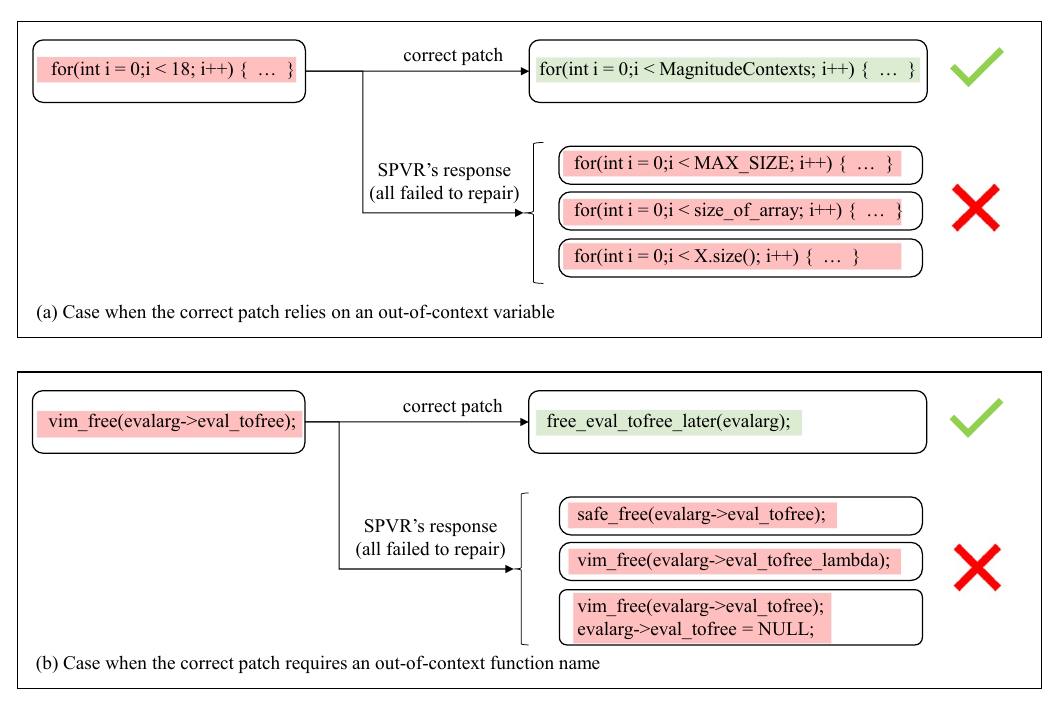}
  \caption{Case study: Cases unable to generate correct patches.}
  \label{case2}
\end{figure}

In contrast to the successful cases discussed above, there are scenarios where \tool is unable to generate the correct patch, and such cases are equally challenging for all other approaches.
These situations typically occur when the ground-truth fix relies on variables or constants that are completely absent from the full context available in the original dataset (e.g., repairing \texttt{for(int i = 0; i < 18; i++)} to \texttt{for(int i = 0; i < MagnitudeContexts; i++)}), or when the required function names are entirely missing from the accessible context (e.g., replacing \texttt{vim\_free(evalarg->eval\_tofree)} with \texttt{free\_eval\_tofree\_later(evalarg)}), shown in Figure~\ref{case2}.

In such cases, LLMs can only propose plausible but non-identical alternatives based on the limited information at hand; however, precisely reproducing the ground-truth fix becomes exceedingly difficult.
This contrast highlights both the strength of \tool when sufficient semantic cues are present and the inherent limitation of LLM-based repair methods when critical information is absent from the original context.

\begin{tcolorbox} 
\parh{Answer to RQ5:} \tool repairs vulnerabilities effectively when the code provides clear semantic cues, such as missing variables or operator misuse. Even when the ground-truth fix relies on external context, \tool leverages available semantics to produce close fixes and maintain strong performance.
\end{tcolorbox}

\section{Threats to Validity}
\label{sec:discussion}



\parh{Real-World Repair Workflows}  
In real-world projects, vulnerability detection and repair workflows often involve executing code and leveraging the expertise of vulnerability specialists. While these mature and complex pipelines are beyond the scope of the current study, improving the foundation of LLM-based vulnerability repair can benefit downstream workflow stages. By enhancing accuracy and efficiency, \tool can streamline the repair process and reduce the cognitive load on human experts. Integrating \tool into existing pipelines and evaluating its impact on real-world projects would provide valuable insights into its practical applicability.

\parh{Applicability to Diverse Programming Languages}  
Although our experiments focus on C/C++ vulnerabilities, the underlying approach can potentially generalize to other programming languages. The MET abstraction and context extraction rules can be adapted to accommodate the unique syntax and semantics of different programming languages. Since \tool leverages Tree-sitter, a universal parser interface supporting over 17 languages, the pipeline can be readily extended, making cross-language application feasible with minimal engineering effort.

\parh{Insights into Prompt Design and Vulnerability Semantics}  
Our study highlights the importance of integrating vulnerability-specific patterns into LLM prompting. Unlike conventional methods that treat repair as a generic sequence-to-sequence task, \tool leverages the MET and CWE information to encode semantic constraints and repair-relevant context. This targeted prompting not only increases the likelihood of generating correct patches but also provides interpretable guidance on vulnerable code regions. Moreover, our approach primarily relies on semantic information and vulnerability patterns rather than superficial semantics. Even when code is poorly formatted, lacks comments, or uses suboptimal naming, \tool can still extract essential semantics and generate effective repairs.

\section{Conclusion}
\label{sec:conclusion}

We introduce \tool, a syntax-to-prompt framework that integrates the Abstract Syntax Tree with CWE IDs to generate targeted, vulnerability-aware repair instructions. Central to \tool is the Minimum Edit Tree, which captures fine-grained structural edits and supports a rule set that extracts repair-relevant context and instantiates precise prompts. Extensive experiments across five Large Language Models demonstrate that \tool markedly improves repair performance, achieving a \textbf{170\%} relative gain in Pass@3 and a \textbf{41\%} gain in CodeBLEU over direct prompting. Ablation and case studies further show that MET-based prompts surface semantic constraints and reduce the search space, enabling correct repairs where generic prompts fail.

Beyond empirical gains, \tool shows that combining static structural analysis with LLM generation yields complementary strengths—precise syntactic cues and flexible code synthesis. 
Being LLM-agnostic, it integrates seamlessly into diverse repair frameworks. Moreover, as the datasets are collected from real-world scenarios with various code quality, the results demonstrate that SPVR remains effective across different code qualities, highlighting its robustness. To further validate this, we conducted additional experiments on a newly collected set of 300 vulnerability cases from the past year, published after the release of most LLMs. The consistent performance on this dataset confirms that SPVR is effective across code with varying quality, reinforcing its practical robustness.
In summary, \tool provides a practical, generalizable approach to vulnerability repair, consistently improving accuracy and offering actionable guidance for real-world software maintenance.

\begin{appendices}

\section{Detailed Information of Check Items}\label{secA1}

\subsection{Extra rules in assignment\_expression:}
\begin{itemize}
    \item Min/max Check:  This rule is used to inspect whether the min/max functions are used in the assignment\_expression. If min/max functions are not present in the MET, MET-related information will include using them and the variables provided in the Possible Variable Use.
    \item Ternary Operator Check:  This rule is used to inspect whether the assignment contains a ternary operator. If ternary operators are not present in the MET, MET-related information will include using one and the variables provided in the Possible Variable Use.
\end{itemize}

\subsection{Extra rules in call\_expression:}
\begin{itemize}
    \item Related Words Check:  This rule is used to inspect function names related to buffers. If a function name contains at least one of [\verb|mem|/\verb|str|/\verb|cpy|] is present in the MET, MET-related information will include adding a sizeof(...) to the variables of the function to avoid potential buffer overflow.
    \item Scope Resolution Operator Check:  This rule is used to inspect whether the call\_expression contains a scope resolution operator. If scope resolution operators are not present in the MET, MET-related information will include using one and the variables provided in the Possible Variable Use.

\end{itemize}

\subsection{Extra rules in declaration:}
\begin{itemize}
    \item Initialization Check: This rule is used to inspect whether all variables are correctly initialized. If the declaration is not initialized in the MET, MET-related information will include initializing it.
    \item Pointer Check: This rule is used to inspect whether the declaration contains a pointer. If pointers are not present in the MET, MET-related information will include checking if the pointer variable is initialized correctly or consider not using pointers.

\end{itemize}

\subsection{Extra rules in function\_definition:}
\begin{itemize}
    \item Static Method Check: This rule is used to inspect whether the function\_definition is static. If not, MET-related information will include reconsidering whether this is a static method.

\end{itemize}

\end{appendices}

\bibliography{ref2}

\end{document}